\documentclass[fleqn,usenatbib]{mnras}

\usepackage{newtxtext,newtxmath}

\usepackage[T1]{fontenc}
\usepackage{ae,aecompl}

\usepackage{graphicx}	
\usepackage{amsmath}	
\usepackage{newtxtext,newtxmath}    
\usepackage{threeparttable, tablefootnote}

\DeclareMathOperator{\sech}{sech}

\title[Bar Buckling and Boxy/Peanut Bulges]{The Effect of Dark Matter Halo Shape on Bar Buckling and Boxy/Peanut Bulges}
\author[A. Kumar et al.]{
Ankit Kumar,$^{1,2}$\thanks{E-mail: ankit4physics@gmail.com (AK)}
Mousumi Das,$^{1}$
and Sandeep Kumar Kataria$^{3}$
\\
$^{1}$Indian Institute of Astrophysics, Bengaluru, 560034, India\\
$^{2}$Joint Astronomy Program, Department of Physics, Indian Institute of Science, Bengaluru, 560012, India \\
$^{3}$School of Physics and Astronomy, Shanghai Jiao Tong University,No.800, Dongchuan Road, Minhang District, Shanghai, China. 
}

\date{Accepted XXX. Received YYY; in original form ZZZ}

\pubyear{2021}

\begin{document}
\label{firstpage}
\pagerange{\pageref{firstpage}--\pageref{lastpage}}
\maketitle

\begin{abstract}
It is well established that bars evolve significantly after they form in galaxy discs, often changing shape both in and out of the disc plane. In some cases they may bend or buckle out of the disc plane resulting in the formation of boxy/peanut/x-shape bulges. In this paper we show that the dark matter halo shape affects bar formation and buckling. We have performed N-body simulations of bar buckling in non-spherical dark matter halos and traced bar evolution for 8~Gyr. We find that bar formation is delayed in oblate halos, resulting in delayed buckling whereas bars form earlier in prolate halos leading to earlier buckling. However, the duration of first buckling remains almost comparable. All the models show two buckling events but the most extreme prolate halo exhibits three distinct buckling features. Bars in prolate halos also show buckling signatures for the longest duration compared to spherical and oblate halos. Since ongoing buckling events are rarely observed, our study suggests that most barred galaxies may have more oblate or spherical halos rather than prolate halos. Our measurement of BPX structures also shows that prolate halos promote bar thickening and disc heating more than oblate and spherical halos.

\end{abstract}




\section{Introduction}
\label{sec:intro} 
The fraction of bars in the observable Universe varies from $30\%$ to $70\%$ \citep{Aguerri.etal.2009, Masters.etal.2011, Diaz-Garcia.etal.2016}. Bars are the major driver of secular evolution in disc galaxies \citep{Weinberg.etal.2007, Athanassoula2013book, Long.etal.2014, Gadotti.etal.2020} and play a crucial role in the re-distribution of the mass and angular momentum among different components of a galaxy \citep{Athanassoula.2002, Athanassoula.2003, Saha.etal.2012, saha.etal.2016, kataria.das.2019}. Studies have shown that during disc evolution bars themselves also evolve, both in length and vertical thickening. The latter can result in the formation of boxy/peanut/x-shape (in short BPX) pseudo-bulges \citep{Friedli1990, Debattista2006, Gadotti2011}.


The origin of the vertical thickening of bars has been studied widely using  numerical simulations \citep{Combes1981, Combes.etal.1990, Raha.etal.1991, Athanassoula.etal.2005}. There are mainly three mechanisms which lead to bar thickening:- (i) bar buckling \citep{Combes.etal.1990, Raha.etal.1991}, (ii) the 2:1 vertical resonance \citep{Quillen.etal.2014}, and (iii) the gradual trapping of stellar orbits into the 2:1 resonance \citep{Sellwood.Gerhard.2020}. Bar buckling is the most violent thickening mechanism during which the bar bends out of the disc plane. The buckling reduces the size and strength of the bar and results in the formation of boxy/peanut bulges \citep{Sellwood.Merritt.1994, Debattista.etal.2005, Martinez-Valpuesta.etal.2004}. In some cases where buckling is slow and less energetic, it can even destroy the bar \citep{Collier.2020}. There are evidences of multiple (recurrent) buckling in numerical simulations \citep{Martinez-Valpuesta.etal.2006, Lokas.2019B}. The primary buckling occurs in the inner part of the bar and persists for less than 1~Gyr; however the secondary buckling take place in the outer region of the bar and remains for few $Gyr$ \citep{Martinez-Valpuesta.etal.2006, Athanassoula2016book}. In this paper we focus only on bar buckling, which is the most rapid process that produces the vertical thickening of bars. 

Some studies suggest that properties of disc, bulge, and gas affect bar buckling. For example in their N-body simulations, \cite{Friedli1990} found that a small asymmetry about the mid plane of the disc accelerates bar buckling. The presence of a classical bulge in the center of galaxies may also prevent the onset of buckling instability in bars \citep{Smirnov.2019}. The warm gas in the galaxies has also been found to decrease the distortion in a bar which is due to buckling \citep{Lokas.2020}. Buckling time and strength remain unaltered in galaxy flybys \citep{Kumar.etal.2021}.

On a large scale, bar formation and evolution are affected by dark matter halo properties. The presence of live dark matter halo supports the bar formation instability, in contrast to rigid dark matter halo that delay onset of bar formation \citep{Athanassoula.2002, Saha.Naab.2013}. There have been several studies of angular momentum re-distribution between bar and live dark matter halo \citep{Sellwood.1980, Athanassoula.2003, Martinez-Valpuesta.etal.2006, Collier.etal.2019}. The detailed study of the effect of halo triaxiality and gas mass fraction on the formation and evolution of the bars is discussed in \cite{Berentzen.etal.2006} and \cite{Athanassoula.etal.2013}. Recently, \cite{Collier.etal.2018} have studied the evolution of bars in rotating and non-spherical live halos. They noticed multiple (=two) buckling events only in prolate halo. The detailed nature of bar buckling in non-spherical halos and the evolution of the buckling induced boxy/peanut bulges are not well explored.

There have been a few attempts to find the signatures of ongoing buckling in the observable Universe. However, the time period of bar buckling is very short and the presence of a central concentration can halt the buckling instability, so it is not easy to detect ongoing bar buckling in galaxies. The first attempt used the bar isophotes and the kinematic signatures of bar buckling derived from simulations to detect buckling events in NGC 3227 and NGC 4569 \citep{Erwin.Debattista.2016}. Recently, \cite{Xiang.etal.2021} have used kinematic signatures to detect ongoing buckling in face-on galaxies.

In this paper we show that the non-spherical nature of halos affects bar buckling significantly and has very important implications for the observations of bars. We vary the halo shape from oblate to prolate, keeping the ratio of halo axes equal in the disc plane, which is a fairly good assumption as shown in recent numerical studies \citep{Bett.etal.2010, Liao.etal.2017}. We have also characterized the properties of the buckling induced boxy/peanut bulges in oblate, spherical and prolate halos.

\section{Simulations And Analysis}
We have simulated isolated disc galaxies with different dark matter halo shapes ranging from oblate to prolate including spherical. All of our model galaxies form a bar and undergo buckling instability. As a result of buckling, the bar thickens and forms a boxy/peanut bulge. To trace the properties of the buckling induced boxy/peanut pseudo-bulge, we have evolved all the models until 8~Gyr. The model setup has been described in detail in earlier papers \citep{kataria.das.2018, Kumar.etal.2021}, and so we describe it only briefly below.

\subsection{Model Galaxies:}
\label{sec:galaxy_model} 
We used a publicly available open source code GALIC \citep{Yurin2014} to generate the initial conditions of our model galaxies. GALIC populates the particles according to the given density distribution and finds the equilibrium solution of the collisionless Boltzmann Equation (CBE) by iteratively changing the initial velocities of the particles. This code makes use of Schwarzschild's method and made-to-measure technique (see \cite{Yurin2014}).

Each of our model galaxies incorporates a stellar disc and a dark matter halo. Since we are interested in bar buckling and the resulting pseudo-bulges, we did not include the bulge component in our models because the presence of a spherical potential in the center of a galaxy slows down bar formation and hence hinders the buckling of bars \citep{kataria.das.2018}. To populate the particles of a spherical dark matter halo, we used the Hernquist density profile \citep{Hernquist1990} defined as,
\begin{equation}
    \rho_{dm}(r)=\frac{M_{dm}}{2\pi}\frac{a}{r(r+a)^3}
    \label{eqn:halo}
\end{equation}
where $M_{dm}$ is the total mass of the dark matter halo and `$a$' is its scale radius and is related to the concentration parameter `$c$' of the NFW halo \citep{NFW1996} with mass $M_{200}=M_{dm}$ by the relation,
\begin{equation}
    a=\frac{r_{200}}{c}\sqrt{2[\ln{(1+c)-\frac{c}{(1+c)}}]},
    \label{eqn:scale_radius}
\end{equation}
where $r_{200}$ is the radius of the NFW halo. It is defined as the radius of the sphere within which the average density is 200 times the critical density of the Universe and $M_{200}$ is the mass within this radius.

The non-spherical halos (oblate and prolate) are generated by linearly distorting the spherical halo along the z-axis perpendicular to the disc plane. Assuming an ellipsoid has $a_{x} = b_{y}$, and $c_{z}$ as its three axes, then the ratio $q=c_{z} / a_{x}$ defines the shape of the halo. An oblate halo has $q<1$, a spherical halo has $q=1$, and a prolate halo has $q>1$. The density profile of such non-spherical halos is given by
\begin{equation}
    \bar \rho_{dm}(R,z,q)=\frac{1}{q} \rho_{dm}(\sqrt{R^{2}+\frac{z^{2}}{q^{2}}})
    \label{eqn:non_spherical_halo}
\end{equation}
where $\rho_{dm}$ is the Hernquist density profile as shown in equation (~\ref{eqn:halo}). This new profiles keeps the total mass of the halo invariant. 

The distribution of particles in the stellar disc is represented by the exponential profile in the radial direction and the $\sech^{2}$ profile in the direction perpendicular to the disc. So the net density profile is given as,
\begin{equation}
   \rho_{d}(R,z)=\frac{M_{d}}{4\pi z_{0} R_{s}^{2}}\exp(-{\frac{R}{R_{s}}}) \sech^{2}(\frac{z}{z_{0}}), 
    \label{eqn:disc}
\end{equation}
where $M_{d}$ is the total disc mass, $z_{0}$ is the disc scale height and $R_{s}$ is the disc scale radius.

\begin{table}
\centering
\caption{Initial parameters of the model galaxies.}
\label{tab:initial_parameters}
\begin{threeparttable}
\begin{tabular}{lr}
\hline
Total mass ($M$) & 6.4 $\times$ $10^{11} M_{\odot}$ \\
\hline
Halo spin parameter($\lambda$) & 0.035 \\
\hline
Halo concentration parameter($c$) & 20 \\
\hline
Disc mass fraction & 0.10 \\
\hline
Disc scale radius ($R_{s}$) & 2.90~kpc \\
\hline
Disc scale height ($z_{0}$) & 0.58~kpc \\
\hline
Halo particles ($N_{Halo}$) & 2.0$\times 10^{6}$ \\
\hline
Disc particles ($N_{Disc}$) & 2.0$\times 10^{6}$ \\
\hline
Total particles ($N_{Total}$) & 4.0$\times 10^{6}$ \\
\hline
\end{tabular}
\end{threeparttable}
\end{table}

\begin{figure*}
    \centering
	\includegraphics[width=\textwidth]{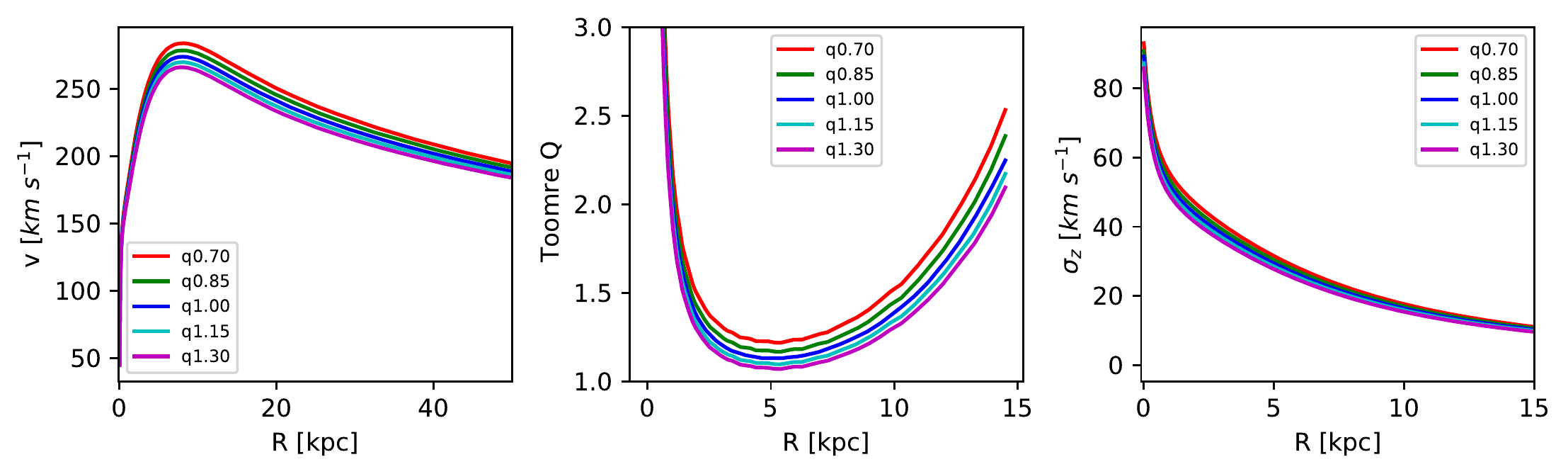}
    \caption{Initial condition of model galaxies. Panels from left to right show the rotation curve, Toomre instability parameter, and vertical velocity dispersion of model disc galaxies respectively. Different colors in the legend represent the models with different the halo shape parameters.}
    \label{fig:init_cond}
\end{figure*}

We have simulated five models of non-spherical dark matter halos in isolated disc galaxies for the halo shape parameter $q \in \{0.70, 0.85, 1.00, 1.15, 1.30\}$. Each of our model galaxies has $2\times 10^{6}$ dark matter particles and $2\times 10^{6}$ stellar particles making a total of $4\times 10^{6}$ particles. For testing purposes, we have also simulated all the models with $2\times 10^{6}$ total particles and found similar results as that for $4\times 10^{6}$ total particles. The total mass of each galaxy is set to $6.4\times 10^{11} M_{\odot}$, where each galaxy contains $90\%$ dark matter mass and $10\%$ stellar mass. Table~\ref{tab:initial_parameters} summarises the initial parameters of our model galaxies which are standard in all the models. The rotation curve $v(R)$, Toomre Q parameter, and the vertical velocity dispersion $\sigma_{z}$ are shown in Fig.~\ref{fig:init_cond}. Since the equilibrium model of the galaxy is generated by iteratively solving CBE for disc and halo, the rotation velocity, Toomre Q, and the velocity dispersion of oblate halos are always a little higher than the respective prolate halos.

After generating initial realizations of model galaxies, we evolved them using the open source code Gadget-2 \citep{Springel2001, Springal2005man, Springel2005} upto 8~Gyr. Gadget-2 is a massively parallelized and adaptive (in both space and in time) code. The gravitational softening for stellar particles and dark matter particles are set to 0.02~kpc and 0.03~kpc respectively. The maximum percentage change in the total angular momentum of the galaxy is well within 0.1$\%$ for all model galaxies throughout the evolution.

We used the shape parameter $'q'$ of the halos for naming the models. For example 'q1.15' represents the initial galaxy model with shape parameter $q=1.15$. However, the shape of the halo may change after evolution of the models. It should be noted that all the quantities discussed here will be in units of the dimensionless Hubble parameter $'h'$ where the Hubble constant is defined as $H_{0}= 100$ $h$ $km$ $s^{-1} Mpc^{-1}$.

\subsection{Analysis:}
\label{sec:analysis}
The bar strength is usually determined using the amplitude of the m=2 Fourier mode relative to m=0 Fourier mode. The amplitude of the $m$th Fourier mode at a cylindrical radius $R$ is given by,
\begin{align}
    A_{m}(R) &= \left| \sum_{j=1}^{N} m_{j} \exp{(i m \phi_{j})} \right|,
    \label{eqn:bar_strenght}
\end{align}
where $m_{j}$ is the mass and $\phi_{j}$ is the azimuth angle of the $j$th particle at a radius $R$, and $N$ is the total number of particles at a radius $R$. The strength of the bar is defined as $A_{bar} = max \left[ \frac{A_{2}}{A_{0}}(R) \right]$.

For the quantification of the buckling instability, we have  adopted the commonly used expressions in the literature \citep{Debattista2006, Xiang.etal.2021}, i.e. the $m=2$ Fourier mode weighted by the vertical velocity
\begin{align}
    A_{buck,v_{z}}(R) &= \left| \frac{\sum_{j=1}^{N} v_{z_j} m_{j} \exp{(2 i \phi_{j})}}{\sum_{j=1}^{N} m_{j}} \right|,
    \label{eqn:buck_strenght_vz}
\end{align}
and the $m=2$ Fourier mode weighted by the vertical height
\begin{align}
    A_{buck,z}(R) &= \left| \frac{\sum_{j=1}^{N} z_{j} m_{j} \exp{(2 i \phi_{j})}}{\sum_{j=1}^{N} m_{j}} \right|.
    \label{eqn:buck_strenght_z}
\end{align}
Equation~\ref{eqn:buck_strenght_vz} is a very useful relation for quantifying buckling in face-on galaxies where stellar kinematic is known while equation~\ref{eqn:buck_strenght_z} quantifies the buckling in edge-on galaxies where vertical stellar morphology is known. We have also used median vertical height ($z_{med}$) for the quantification of buckling. The evolution of the boxy/peanut/x-shape structure is traced using the root mean square vertical height ($A_{BPX} = z_{rms}$). For all these calculations, we have considered only those stellar particles which lie above and below the mid plane within 10 disc scale height ($10 z_{0}$). This choice reduces the discreteness noise from the measurement when some particles reach higher vertical distances.

\section{Results}
\label{sec:results}
We have evolved isolated disc galaxy models with varying dark matter halo shapes ranging from q=0.70 (oblate) to q=1.30 (prolate). All the models are bar unstable and show bar buckling instability after reaching maximum bar strength. In the following subsections, we  discuss how bar buckling is affected by halos shape and how it affects the final product, the boxy/peanut pseudo-bulge.

\subsection{Dependence of bar formation time and bar strength on halo oblateness:}
\label{sec:bar_strength}
\begin{figure}
    \centering
	\includegraphics[width=\columnwidth]{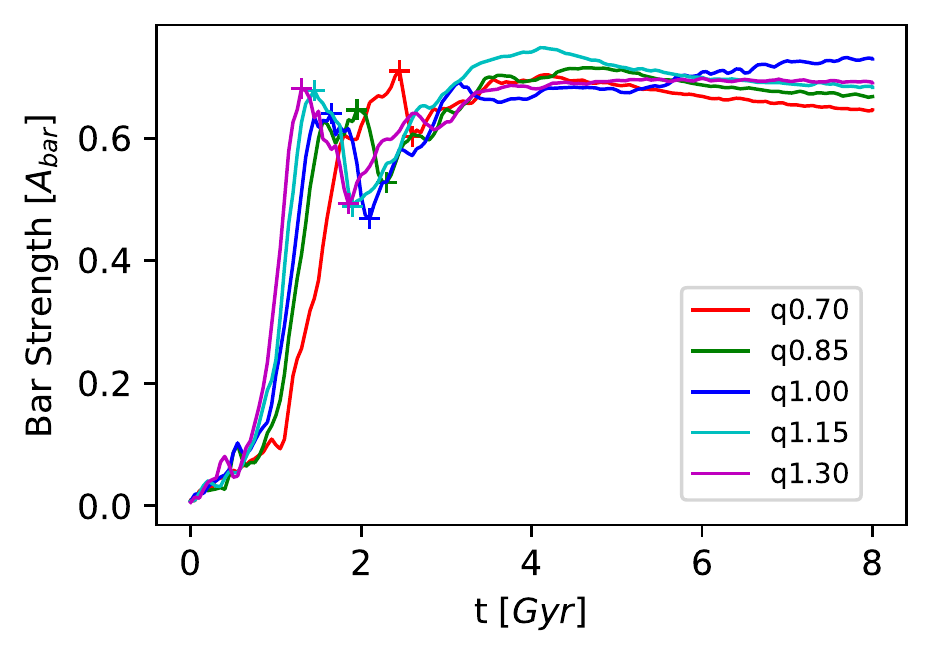}
    \caption{Evolution of bar strength in non-spherical dark matter halos with different shape parameters $'q'$. The positions of bar formation and re-growth after buckling are marked with '+' symbol.}
    \label{fig:bar_strength}
\end{figure}

In Fig.~\ref{fig:bar_strength}, we have shown the time evolution of the bar strength in our model galaxies. We have smoothed the bar strength with the Savitzky–Golay filter \citep{Savgol.1964} using window size = 5, and polynomial order = 3 for the filter parameters. The main motive for using this filter was to remove the noise and reveal the global evolution of the bar instability. All the models show a quick rise in bar strength. After reaching maximum strength, the bar strength decreases a little and again starts increasing. Later in this section we will show that this drop in bar strength is the result of bar buckling (or bending) event which weakens the bar. Finally, the strength of the bar saturates in all the models.

The effect of non-spherical dark matter halos on the bar formation is clearly visible. Prolate halos promote early bar formation, whereas oblate halos delay bar formation (where we consider the bar formation time to be the time from the beginning of simulation to the peak bar strength just before the decrease due to bar buckling). If we consider the galaxy with a spherical halo as the control model, oblate halos delay bar formation by more amount of time compared to the corresponding time taken for prolate halos to promote early bar formation. 

At the time of bar formation, the amplitude of bar strength is always higher in non-spherical halos. The decrease in bar amplitude due to buckling is highest for prolate halos and lowest for oblate halos. The change in bar amplitude is 0.19 for the $q=1.30$ model and 0.11 for the $q=0.70$ model. The difference between the bar formation time and the re-growth time (by re-growth time, we mean the time when bar starts recovering from the weakening caused due to buckling) is longer for prolate halos as compared to oblate halos. This difference is 0.55~Gyr for the $q=1.30$ model halo and 0.15~Gyr for the $q=0.70$ model.

\subsection{Bar buckling in oblate and prolate halos:}
\label{sec:bar_buckling}
\begin{figure}
    \centering
	\includegraphics[width=\columnwidth]{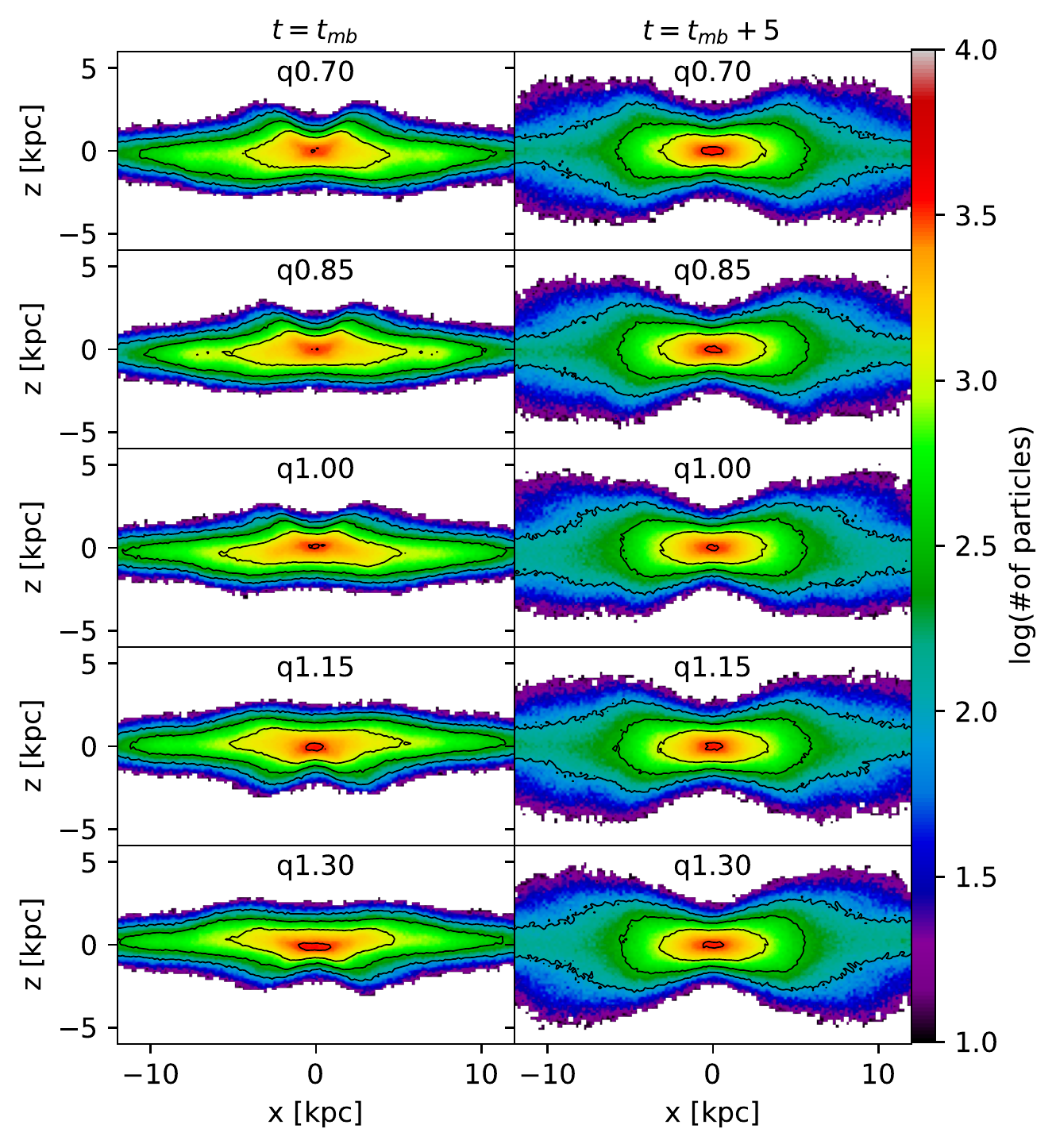}
    \caption{Bar buckling in non-spherical dark matter halos along with iso-density contours. Left column shows the side-on view (perpendicular to bar) of the disc at the time of maximum buckling amplitude ($t_{mb}$) and right column shows the side-on view after 5~Gyr of maximum buckling. }
    \label{fig:bar_buckling_qual}
\end{figure}
All of our model galaxies go though the buckling event during which the bar bends out of the disc plane and and becomes overall weaker. To verify that  the buckling instability takes place, we have shown the edge-on view of the bars in the left column of Fig.~\ref{fig:bar_buckling_qual} at the time of maximum buckling/bending amplitude (hereafter, we use $t_{mb}$ to represent this time) calculated using equation~\ref{eqn:buck_strenght_vz}. The right column of this figure shows the buckling induced boxy/peanut/x-shape pseudo-bulge after 5~Gyr of the maximum buckling amplitude ($t_{mb}$+5~Gyr). The direction of the bending depends on the mass asymmetry around the mid plane of the disc before buckling, and so the vertical direction can vary from model to model \citep{Friedli1990}.

\begin{figure*}
    \centering
	\includegraphics[width=\textwidth]{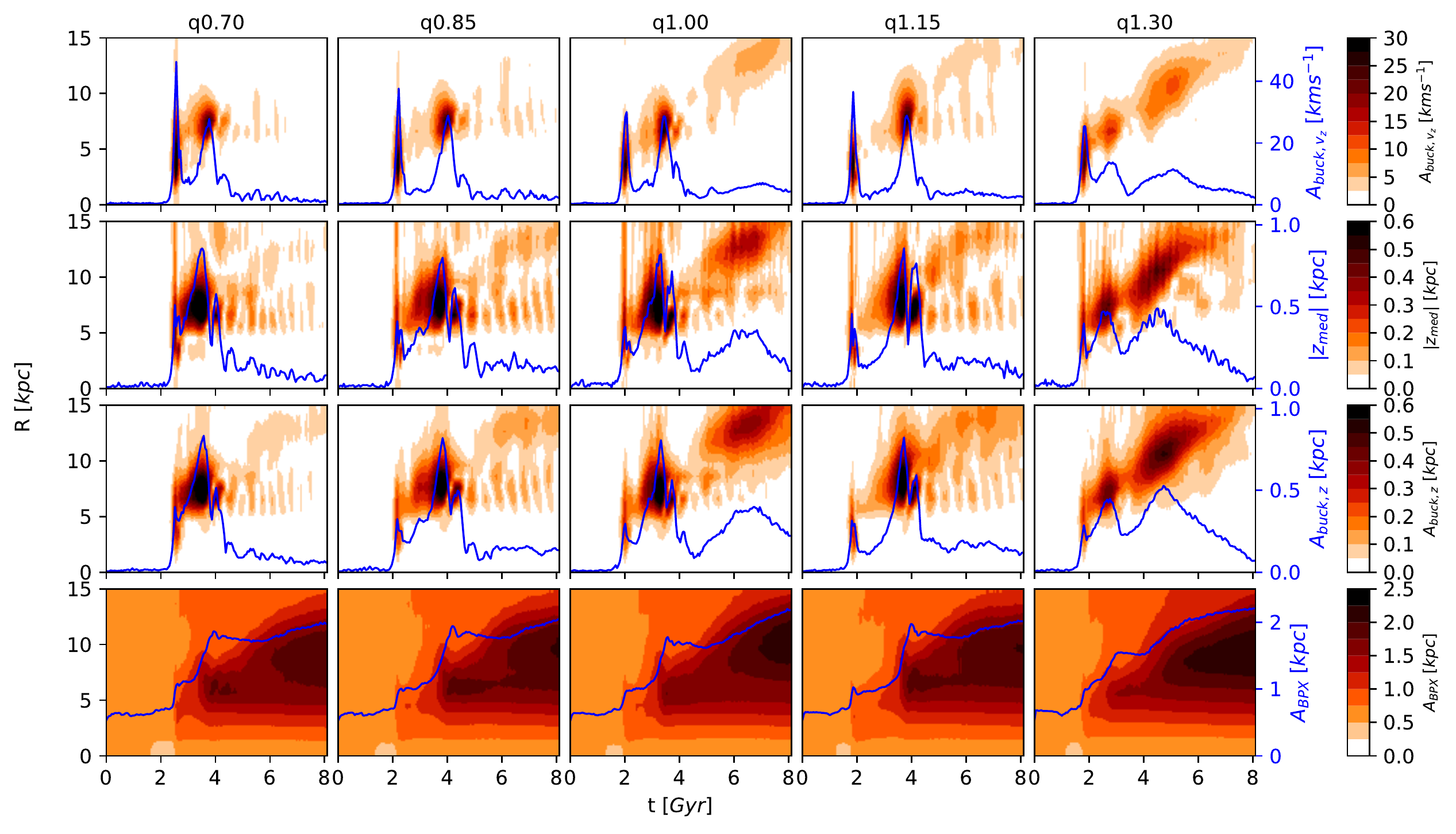}
    \caption{Quantification of bar buckling and resultant BPX structure in non-spherical dark matter halos. First row panels show the buckling amplitude using stellar kinematics while second and third rows panels represent buckling amplitude using stellar morphology. Last row panels show BPX strength. Each column shows a different galaxy model. The temporal variation of maximum strength is shown by over-plotted blue curve.}
    \label{fig:bar_buckling_quant}
\end{figure*}
In Fig.~\ref{fig:bar_buckling_quant}, we have shown the color coded radial and temporal distribution of buckling amplitude (panels in top three rows) and boxy/peanut strength (panels in bottom row) in oblate, spherical and prolate halos. We have also over-plotted each panel with the temporal distribution of the maximum strength. Each column of this figure represents a model galaxy as shown at the top. The panels in the first row show the kinematic amplitude of bar buckling events while the second and third rows quantify morphological asymmetry caused in the bar due to buckling. Here all the high amplitude (i.e. distinct and measurable) peaks correspond to a buckling event during evolution. The fourth row at the bottom demonstrate the strength of boxy/peanut/x-shape structure induced by bar buckling.

As, in Fig.~\ref{fig:bar_strength}, we see a variation in the bar formation time with halo shape. The time of the bar buckling also varies with halo shape. From the first peaks of blue curves in the top row of Fig.~\ref{fig:bar_buckling_quant}, One can notice that the buckling timescale increases with decreasing halo shape parameter $q$. Prolate halos buckle earlier and oblate halos buckle later with respect to spherical halo. The kinematic signature of buckling indicates that oblate halos start buckling just after bar formation, whereas their prolate counterparts take some time to buckle after bar formation. On the other hand, the first buckling event attains its maximum amplitude just before bar re-growth time. These two effects result in closely similar buckling period during first buckling as can be interpreted from the width of the first peak in all the panels of the first row.

\begin{figure}
    \centering
	\includegraphics[width=\columnwidth]{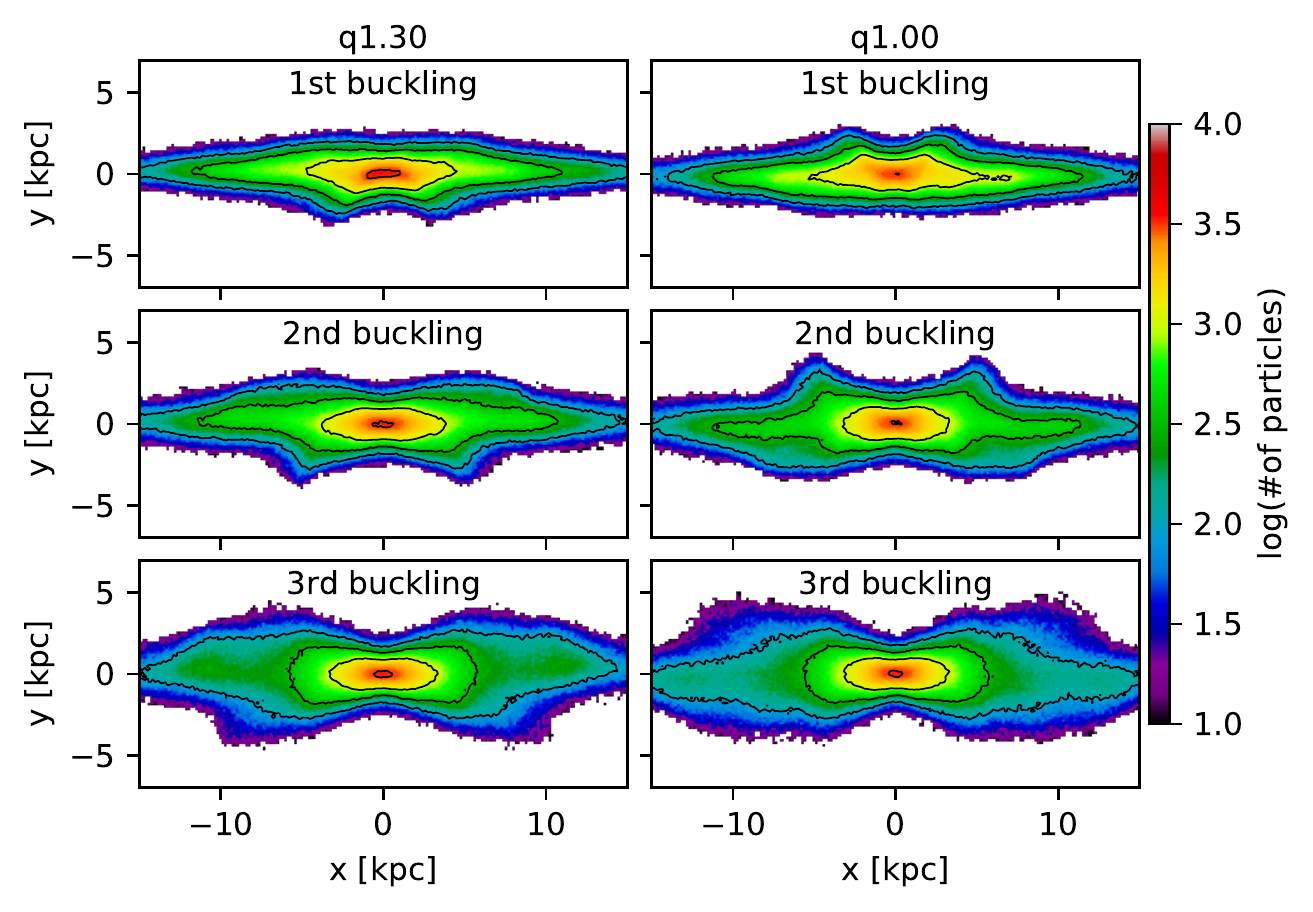}
    \caption{Triple bar buckling events in q=1.30 halo along with iso-density contour compared with the q=1.00 halo. Each panel is shown at the time of distinct peaks in $A_{buck,v_{z}}$ as shown by blue curves in the first row of Fig.~\ref{fig:bar_buckling_quant}. A clear bending of the iso-density contour can be noticed in the outer bar of the prolate halo model during 3rd buckling event.}
    \label{fig:triple_buckling}
\end{figure}

By comparing the panels of the first row in Fig.~\ref{fig:bar_buckling_quant}, one can clearly notice that the kinematic signature of first buckling is always higher in amplitude than the succeeding buckling events. However, the morphological signatures have higher amplitude during the second buckling event as can be seen in the second and third row panels of the figure. When moving from oblate to prolate halos in any of the first three rows, we can spot a clear difference of increasing buckling signature at outer radial positions. This implies that prolate halos help the bar to buckle in the outer parts of the bar while oblate halos suppress the buckling at outer edges. At the outer edge of the bar, the buckling strength of the q1.15 model is slightly weaker than the q1.00 model. But it is strongest for the q1.30 model which indicates that the small deviation from spherical shape towards prolateness decreases the buckling amplitude but more deviation increases it. Prolate halo for q=1.30 shows an explicitly distinct buckling strength at the outer edges of the bar in both kinematic and morphological quantification which qualifies it to be a third buckling event. The spherical halo with q=1.00 also shows a weak signature of the third buckling event. In Fig.~\ref{fig:triple_buckling}, we have shown the third buckling event in our most prolate halo and compared it with the spherical model. During the third buckling event, the iso-density contours in the prolate halo show clear noticeable distortion in the disc while the corresponding distortion in the spherical halo is very weak. Hence, the most prolate halo in our set of models has a halo shape that promotes the onset of three buckling events in the bar.

Seeing the widths of the peaks in each panel of the top three rows of Fig.~\ref{fig:bar_buckling_quant}, one can easily interpret that the duration of a preceding buckling is always smaller than the succeeding one. Since, the signatures of first buckling are short lived so, it has a lower probability to be detected in observations of galaxies. Instead it is the second buckling event which has a higher probability of being detected as an ongoing buckling event. In our models, the prolate halo of q=1.30 show a remarkable third episode of buckling, as can be seen in the last column of the figure. This suggests that in observations of galaxies, buckling events are more likely to be associated with prolate halos rather than oblate or spherical halos.

\subsection{Effect of halo shape on the boxy/peanut bulge:}
\label{sec:bpx_bulge}
The last row of Fig.~\ref{fig:bar_buckling_quant} shows the evolution of the boxy/peanut pseudo-bulge in distorted halos. There is one to one correlation between the bar buckling event and the steep (or sudden) rise in BPX strength. After each buckling event, the bar quickly gets thicker as can be seen in each panel of the bottom row. There is no significant difference in the inner bar BPX strength for different halos. But, the outer part of the bar in prolate halos gets more thicker than the oblate halos. The same result can be interpreted from the closer look at iso-density contours in the right column of Fig.~\ref{fig:bar_buckling_qual}. It shows that oblate halos try to restrict bar/disc against vertical thickening, whereas prolate halos promote vertical heating by the mean of continuous buckling.

\begin{figure}
    \centering
	\includegraphics[width=\columnwidth]{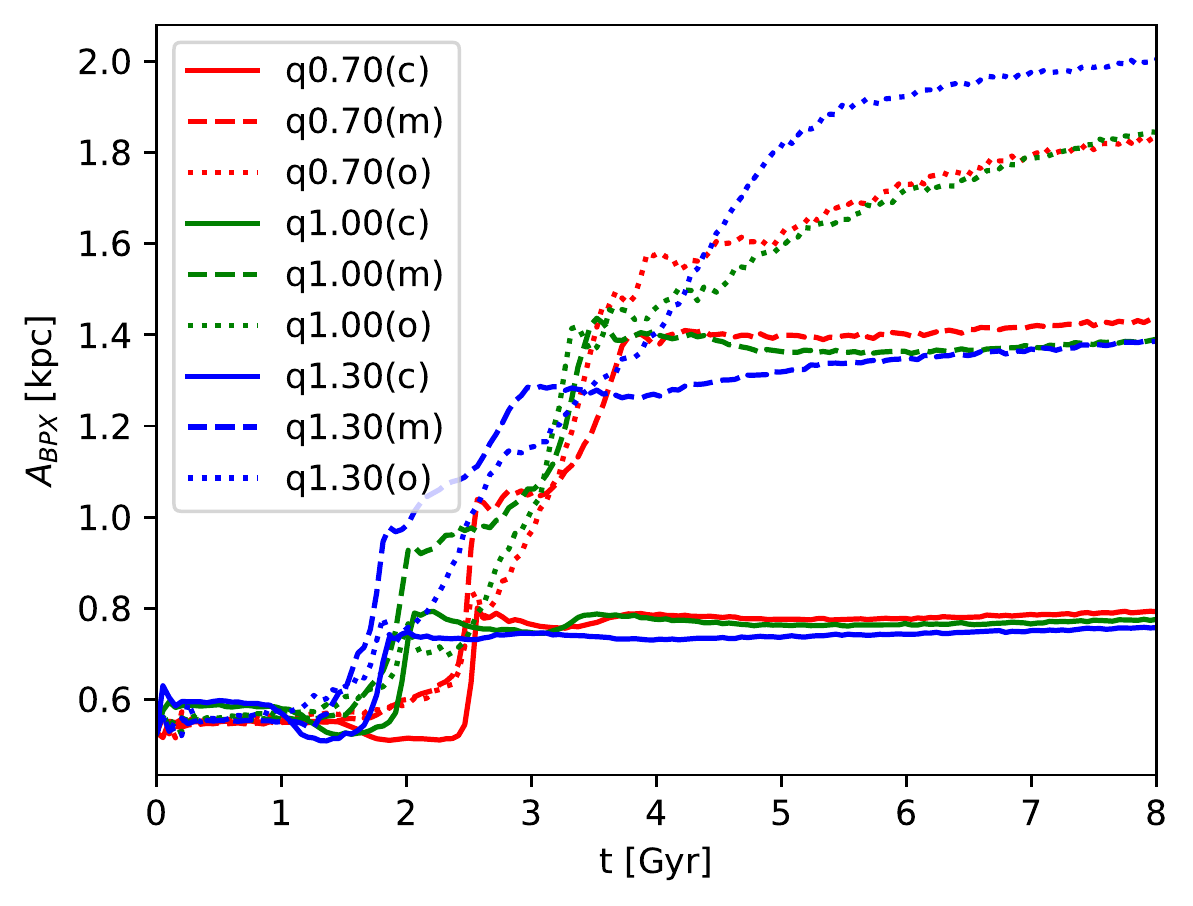}
    \caption{ Evolution of boxy/peanut/x-shape (BPX) structure in three different regions of the disc. For the sake of better visualization, only three extreme models (q0.70, q1.00, q1.30) are shown with different colors. Here, solid curves show central region $R \in [0,R_{s})$, dashed curves represent middle region $R \in [R_{s},2R_{s})$, and dotted curves demonstrate outer region $R \in [2R_{s},3R_{s})$ as marked with symbols 'c', 'm', and 'o' respectively in the legend.}
    \label{fig:bpx_strength}
\end{figure}

The blue curves in the bottom row panels of Fig.~\ref{fig:bar_buckling_quant} represent the time evolution of the maximum BPX strength in the shown region of the disc. Thus, for the sake of better visualization, we have also shown the strength of BPX pseudobulges in three different parts of the disc for the three extreme models q0.70, q1.00 and q1.30 in Fig.~\ref{fig:bpx_strength}. In the figure the solid curves show the central region $R \in [0,1R_{s})$, dashed curves represent middle region $R \in [1R_{s},2R_{s})$, and dotted curves demonstrate outer region $R \in [2R_{s},3R_{s})$. These regions are marked with symbols 'c', 'm', and 'o' respectively in the legend of the figure. The solid curves of Fig.~\ref{fig:bpx_strength} clearly show that the central part of the bar that lies within disk scale radius thickens least and the amplitude is fixed in all the models. Similarly, the middle part of the bar shows nearly equal BPX amplitude in all the models after evolution but its magnitude is higher than that of the central part of the bar. The outer bar region shows more thickening than the central and middle regions for all the models. But the outer region also shows the maximum thickening for the prolate halo as compared to the spherical and oblate halos.

\section{Conclusions and Discussion}
\label{sec:conclusions}
We have performed N-body simulations of bar buckling in isolated disc galaxies with different dark matter halo shapes ranging from oblate to prolate. The Fourier analysis techniques are used for the quantification of the bar and buckling strength. The main findings of this work are listed below.

(i) Oblate halos delay the bar formation and provide support to the bar against the weakening due to buckling instability, whereas prolate halos helps in early bar formation and this bar weakens more during buckling. But, the re-growth of the bar does not leave any signature of halo shape in bar amplitude. Bar strength become same in both types of halos after recovering from first buckling.

(ii) Though the bar in the prolate halo takes a longer time to re-grow than the bar in an oblate halo, the time period of first buckling event remains closely similar. This is because prolate halos take extra time to buckle after bar formation but oblate halos buckle just after bar formation. On the other hand, introducing small prolateness in the spherical halo suppresses the third buckling event that is seen in the spherical halo but, increasing more prolateness promotes the third buckling event while oblate halos always suppress third bar buckling. In our models, all the oblate halos show only two distinct buckling events. However, the most prolate halo in our set of simulations, q=1.30, displays three noticeable well defined buckling events. 

(iii) The buckling induced boxy/peanut/x-shape structure (or BPX bulge) remains closely similar in the inner part of the bar irrespective of the dark matter halo shape. But, outer edge of the bar thickens more in prolate dark matter halos as compared to oblate and spherical dark matter halos. Prolate halos make bar/disc more thicker than the oblate and spherical dark matter halos.

Oblate halos provide more gravitational potential near the disc plane as compared to prolate halos (see Fig.~\ref{fig:init_cond}). As a result, the rotation velocity, Toomre instability parameter, and vertical dispersion of the disc particles increases and bar formation is delayed in oblate halos. In our set of simulations, the most prolate halo with shape parameter q=1.30 shows the noticeable third buckling event and the spherical model shows only a very weak third buckling event. All the oblate halos and the halo with small prolateness suppress the signature of third buckling event. On the other hand, the timescale of first and second buckling events are very small as compared to the third one. This itself explains why the detection of ongoing buckling events in the observations of nearby galaxies is such a small fraction of all barred galaxies. Till date only 8 buckling events have been observed (see Table 2 in \cite{Xiang.etal.2021}) using various existing methodologies.

Our simulations do not account for the presence of gas, classical bulge and interactions with other galaxies. All these factors also affect the bar formation and buckling of the bar. The presence of gas, and classical bulge prevent bar formation whereas flyby interactions have been seen to produce bars \citep{Lokas.etal.2018, Smirnov.2019, Lokas.2020}. Studying the effect of various components of a galaxy on bar buckling and the galaxy environment, may provide constraints on the shapes of dark matter halos. \cite{O'Brien.etal.2010} and many previous studies suggest that dark matter halos shape parameter lies in the range of q=0.1 to q=1.4. A detailed statistical study of halo shapes in observations of galaxies and state of art cosmological simulations may help improve our understanding of bar buckling and the formation of boxy/peanut/x-shape structures.


\section*{ACKNOWLEDGEMENTS}
Authors would like to thank the anonymous reviewer for the constructive comments and suggestions which improved the quality of the paper. We are grateful to high performance computing facility `NOVA' at Indian Institute of Astrophysics, Bengaluru, India where we ran all our simulations. We thank Denis Yurin and Volker Springel for providing GalIC and Gadget-2 code that we have used for this study. This study also made use of NumPy \citep{numpy2020}, Matplotlib \citep{matplotlib2007}, and Astropy \citep{astropy2018} packages. MD acknowledges the support of Science and Engineering Research Board (SERB) MATRICS grant MTR/2020/000266 for this research. 

\section*{DATA AVAILABILITY}
The data generated in this research will be shared on reasonable request to the corresponding author.




\bibliographystyle{mnras}
\bibliography{Kumar} 

\begin{thebibliography}{}
\makeatletter
\relax
\def\mn@urlcharsother{\let\do\@makeother \do\$\do\&\do\#\do\^\do\_\do\%\do\~}
\def\mn@doi{\begingroup\mn@urlcharsother \@ifnextchar [ {\mn@doi@}
  {\mn@doi@[]}}
\def\mn@doi@[#1]#2{\def\@tempa{#1}\ifx\@tempa\@empty \href
  {http://dx.doi.org/#2} {doi:#2}\else \href {http://dx.doi.org/#2} {#1}\fi
  \endgroup}
\def\mn@eprint#1#2{\mn@eprint@#1:#2::\@nil}
\def\mn@eprint@arXiv#1{\href {http://arxiv.org/abs/#1} {{\tt arXiv:#1}}}
\def\mn@eprint@dblp#1{\href {http://dblp.uni-trier.de/rec/bibtex/#1.xml}
  {dblp:#1}}
\def\mn@eprint@#1:#2:#3:#4\@nil{\def\@tempa {#1}\def\@tempb {#2}\def\@tempc
  {#3}\ifx \@tempc \@empty \let \@tempc \@tempb \let \@tempb \@tempa \fi \ifx
  \@tempb \@empty \def\@tempb {arXiv}\fi \@ifundefined
  {mn@eprint@\@tempb}{\@tempb:\@tempc}{\expandafter \expandafter \csname
  mn@eprint@\@tempb\endcsname \expandafter{\@tempc}}}

\bibitem[\protect\citeauthoryear{{Aguerri}, {M{\'e}ndez-Abreu}  \&
  {Corsini}}{{Aguerri} et~al.}{2009}]{Aguerri.etal.2009}
{Aguerri} J.~A.~L.,  {M{\'e}ndez-Abreu} J.,   {Corsini} E.~M.,  2009, \mn@doi
  [\aap] {10.1051/0004-6361:200810931}, \href
  {https://ui.adsabs.harvard.edu/abs/2009A&A...495..491A} {495, 491}

\bibitem[\protect\citeauthoryear{{Astropy Collaboration} et~al.,}{{Astropy
  Collaboration} et~al.}{2018}]{astropy2018}
{Astropy Collaboration} et~al., 2018, \mn@doi [\aj] {10.3847/1538-3881/aabc4f},
  \href {https://ui.adsabs.harvard.edu/abs/2018AJ....156..123A} {156, 123}

\bibitem[\protect\citeauthoryear{{Athanassoula}}{{Athanassoula}}{2002}]{Athanassoula.2002}
{Athanassoula} E.,  2002, \mn@doi [\apjl] {10.1086/340784}, \href
  {https://ui.adsabs.harvard.edu/abs/2002ApJ...569L..83A} {569, L83}

\bibitem[\protect\citeauthoryear{{Athanassoula}}{{Athanassoula}}{2003}]{Athanassoula.2003}
{Athanassoula} E.,  2003, \mn@doi [\mnras] {10.1046/j.1365-8711.2003.06473.x},
  \href {https://ui.adsabs.harvard.edu/abs/2003MNRAS.341.1179A} {341, 1179}

\bibitem[\protect\citeauthoryear{{Athanassoula}}{{Athanassoula}}{2013}]{Athanassoula2013book}
{Athanassoula} E.,  2013, {Bars and secular evolution in disk galaxies:
  Theoretical input}.
p.~305

\bibitem[\protect\citeauthoryear{{Athanassoula}}{{Athanassoula}}{2016}]{Athanassoula2016book}
{Athanassoula} E.,  2016, {Boxy/Peanut/X Bulges, Barlenses and the Thick Part
  of Galactic Bars: What Are They and How Did They Form?}.
p.~391, \mn@doi{10.1007/978-3-319-19378-6\_14}

\bibitem[\protect\citeauthoryear{{Athanassoula}, {Machado}  \&
  {Rodionov}}{{Athanassoula} et~al.}{2013}]{Athanassoula.etal.2013}
{Athanassoula} E.,  {Machado} R. E.~G.,   {Rodionov} S.~A.,  2013, \mn@doi
  [\mnras] {10.1093/mnras/sts452}, \href
  {https://ui.adsabs.harvard.edu/abs/2013MNRAS.429.1949A} {429, 1949}

\bibitem[\protect\citeauthoryear{{Berentzen}, {Shlosman}  \&
  {Jogee}}{{Berentzen} et~al.}{2006}]{Berentzen.etal.2006}
{Berentzen} I.,  {Shlosman} I.,   {Jogee} S.,  2006, \mn@doi [\apj]
  {10.1086/498493}, \href
  {https://ui.adsabs.harvard.edu/abs/2006ApJ...637..582B} {637, 582}

\bibitem[\protect\citeauthoryear{{Bett}, {Eke}, {Frenk}, {Jenkins}  \&
  {Okamoto}}{{Bett} et~al.}{2010}]{Bett.etal.2010}
{Bett} P.,  {Eke} V.,  {Frenk} C.~S.,  {Jenkins} A.,   {Okamoto} T.,  2010,
  \mn@doi [\mnras] {10.1111/j.1365-2966.2010.16368.x}, \href
  {https://ui.adsabs.harvard.edu/abs/2010MNRAS.404.1137B} {404, 1137}

\bibitem[\protect\citeauthoryear{{Bureau} \& {Athanassoula}}{{Bureau} \&
  {Athanassoula}}{2005}]{Athanassoula.etal.2005}
{Bureau} M.,  {Athanassoula} E.,  2005, \mn@doi [\apj] {10.1086/430056}, \href
  {https://ui.adsabs.harvard.edu/abs/2005ApJ...626..159B} {626, 159}

\bibitem[\protect\citeauthoryear{{Collier}}{{Collier}}{2020}]{Collier.2020}
{Collier} A.,  2020, \mn@doi [\mnras] {10.1093/mnras/stz3625}, \href
  {https://ui.adsabs.harvard.edu/abs/2020MNRAS.492.2241C} {492, 2241}

\bibitem[\protect\citeauthoryear{{Collier}, {Shlosman}  \& {Heller}}{{Collier}
  et~al.}{2018}]{Collier.etal.2018}
{Collier} A.,  {Shlosman} I.,   {Heller} C.,  2018, \mn@doi [\mnras]
  {10.1093/mnras/sty270}, \href
  {https://ui.adsabs.harvard.edu/abs/2018MNRAS.476.1331C} {476, 1331}

\bibitem[\protect\citeauthoryear{{Collier}, {Shlosman}  \& {Heller}}{{Collier}
  et~al.}{2019}]{Collier.etal.2019}
{Collier} A.,  {Shlosman} I.,   {Heller} C.,  2019, \mn@doi [\mnras]
  {10.1093/mnras/stz2144}, \href
  {https://ui.adsabs.harvard.edu/abs/2019MNRAS.488.5788C} {488, 5788}

\bibitem[\protect\citeauthoryear{{Combes} \& {Sanders}}{{Combes} \&
  {Sanders}}{1981}]{Combes1981}
{Combes} F.,  {Sanders} R.~H.,  1981, \aap, \href
  {https://ui.adsabs.harvard.edu/abs/1981A&A....96..164C} {96, 164}

\bibitem[\protect\citeauthoryear{{Combes}, {Debbasch}, {Friedli}  \&
  {Pfenniger}}{{Combes} et~al.}{1990}]{Combes.etal.1990}
{Combes} F.,  {Debbasch} F.,  {Friedli} D.,   {Pfenniger} D.,  1990, \aap,
  \href {https://ui.adsabs.harvard.edu/abs/1990A&A...233...82C} {233, 82}

\bibitem[\protect\citeauthoryear{{Debattista}, {Carollo}, {Mayer}  \&
  {Moore}}{{Debattista} et~al.}{2005}]{Debattista.etal.2005}
{Debattista} V.~P.,  {Carollo} C.~M.,  {Mayer} L.,   {Moore} B.,  2005, \mn@doi
  [\apj] {10.1086/431292}, \href
  {https://ui.adsabs.harvard.edu/abs/2005ApJ...628..678D} {628, 678}

\bibitem[\protect\citeauthoryear{{Debattista}, {Mayer}, {Carollo}, {Moore},
  {Wadsley}  \& {Quinn}}{{Debattista} et~al.}{2006}]{Debattista2006}
{Debattista} V.~P.,  {Mayer} L.,  {Carollo} C.~M.,  {Moore} B.,  {Wadsley} J.,
   {Quinn} T.,  2006, \mn@doi [\apj] {10.1086/504147}, \href
  {https://ui.adsabs.harvard.edu/abs/2006ApJ...645..209D} {645, 209}

\bibitem[\protect\citeauthoryear{{D{\'\i}az-Garc{\'\i}a}, {Salo}, {Laurikainen}
   \& {Herrera-Endoqui}}{{D{\'\i}az-Garc{\'\i}a}
  et~al.}{2016}]{Diaz-Garcia.etal.2016}
{D{\'\i}az-Garc{\'\i}a} S.,  {Salo} H.,  {Laurikainen} E.,   {Herrera-Endoqui}
  M.,  2016, \mn@doi [\aap] {10.1051/0004-6361/201526161}, \href
  {https://ui.adsabs.harvard.edu/abs/2016A&A...587A.160D} {587, A160}

\bibitem[\protect\citeauthoryear{{Erwin} \& {Debattista}}{{Erwin} \&
  {Debattista}}{2016}]{Erwin.Debattista.2016}
{Erwin} P.,  {Debattista} V.~P.,  2016, \mn@doi [\apjl]
  {10.3847/2041-8205/825/2/L30}, \href
  {https://ui.adsabs.harvard.edu/abs/2016ApJ...825L..30E} {825, L30}

\bibitem[\protect\citeauthoryear{{Friedli} \& {Pfenniger}}{{Friedli} \&
  {Pfenniger}}{1990}]{Friedli1990}
{Friedli} D.,  {Pfenniger} D.,  1990, in European Southern Observatory
  Conference and Workshop Proceedings. p.~265

\bibitem[\protect\citeauthoryear{{Gadotti}}{{Gadotti}}{2011}]{Gadotti2011}
{Gadotti} D.~A.,  2011, Memorie della Societa Astronomica Italiana Supplementi,
  \href {https://ui.adsabs.harvard.edu/abs/2011MSAIS..18...69G} {18, 69}

\bibitem[\protect\citeauthoryear{{Gadotti} et~al.,}{{Gadotti}
  et~al.}{2020}]{Gadotti.etal.2020}
{Gadotti} D.~A.,  et~al., 2020, \mn@doi [\aap] {10.1051/0004-6361/202038448},
  \href {https://ui.adsabs.harvard.edu/abs/2020A&A...643A..14G} {643, A14}

\bibitem[\protect\citeauthoryear{Harris et~al.,}{Harris
  et~al.}{2020}]{numpy2020}
Harris C.~R.,  et~al., 2020, \mn@doi [Nature] {10.1038/s41586-020-2649-2}, 585,
  357–362

\bibitem[\protect\citeauthoryear{{Hernquist}}{{Hernquist}}{1990}]{Hernquist1990}
{Hernquist} L.,  1990, \mn@doi [\apj] {10.1086/168845}, \href
  {https://ui.adsabs.harvard.edu/abs/1990ApJ...356..359H} {356, 359}

\bibitem[\protect\citeauthoryear{{Hunter}}{{Hunter}}{2007}]{matplotlib2007}
{Hunter} J.~D.,  2007, \mn@doi [Computing in Science Engineering]
  {10.1109/MCSE.2007.55}, 9, 90

\bibitem[\protect\citeauthoryear{{Kataria} \& {Das}}{{Kataria} \&
  {Das}}{2018}]{kataria.das.2018}
{Kataria} S.~K.,  {Das} M.,  2018, \mn@doi [\mnras] {10.1093/mnras/stx3279},
  \href {https://ui.adsabs.harvard.edu/abs/2018MNRAS.475.1653K} {475, 1653}

\bibitem[\protect\citeauthoryear{{Kataria} \& {Das}}{{Kataria} \&
  {Das}}{2019}]{kataria.das.2019}
{Kataria} S.~K.,  {Das} M.,  2019, \mn@doi [\apj] {10.3847/1538-4357/ab48f7},
  \href {https://ui.adsabs.harvard.edu/abs/2019ApJ...886...43K} {886, 43}

\bibitem[\protect\citeauthoryear{{Kumar}, {Das}  \& {Kataria}}{{Kumar}
  et~al.}{2021}]{Kumar.etal.2021}
{Kumar} A.,  {Das} M.,   {Kataria} S.~K.,  2021, \mn@doi [\mnras]
  {10.1093/mnras/stab1742}, \href
  {https://ui.adsabs.harvard.edu/abs/2021MNRAS.506...98K} {506, 98}

\bibitem[\protect\citeauthoryear{{Liao}, {Chen}  \& {Chu}}{{Liao}
  et~al.}{2017}]{Liao.etal.2017}
{Liao} S.,  {Chen} J.,   {Chu} M.~C.,  2017, \mn@doi [\apj]
  {10.3847/1538-4357/aa79fb}, \href
  {https://ui.adsabs.harvard.edu/abs/2017ApJ...844...86L} {844, 86}

\bibitem[\protect\citeauthoryear{{{\L}okas}}{{{\L}okas}}{2018}]{Lokas.etal.2018}
{{\L}okas} E.~L.,  2018, \mn@doi [\apj] {10.3847/1538-4357/aab4ff}, \href
  {https://ui.adsabs.harvard.edu/abs/2018ApJ...857....6L} {857, 6}

\bibitem[\protect\citeauthoryear{{{\L}okas}}{{{\L}okas}}{2019}]{Lokas.2019B}
{{\L}okas} E.~L.,  2019, \mn@doi [\aap] {10.1051/0004-6361/201936056}, \href
  {https://ui.adsabs.harvard.edu/abs/2019A&A...629A..52L} {629, A52}

\bibitem[\protect\citeauthoryear{{{\L}okas}}{{{\L}okas}}{2020}]{Lokas.2020}
{{\L}okas} E.~L.,  2020, \mn@doi [\aap] {10.1051/0004-6361/201937165}, \href
  {https://ui.adsabs.harvard.edu/abs/2020A&A...634A.122L} {634, A122}

\bibitem[\protect\citeauthoryear{{Long}, {Shlosman}  \& {Heller}}{{Long}
  et~al.}{2014}]{Long.etal.2014}
{Long} S.,  {Shlosman} I.,   {Heller} C.,  2014, \mn@doi [\apjl]
  {10.1088/2041-8205/783/1/L18}, \href
  {https://ui.adsabs.harvard.edu/abs/2014ApJ...783L..18L} {783, L18}

\bibitem[\protect\citeauthoryear{{Martinez-Valpuesta} \&
  {Shlosman}}{{Martinez-Valpuesta} \&
  {Shlosman}}{2004}]{Martinez-Valpuesta.etal.2004}
{Martinez-Valpuesta} I.,  {Shlosman} I.,  2004, \mn@doi [\apjl]
  {10.1086/424876}, \href
  {https://ui.adsabs.harvard.edu/abs/2004ApJ...613L..29M} {613, L29}

\bibitem[\protect\citeauthoryear{{Martinez-Valpuesta}, {Shlosman}  \&
  {Heller}}{{Martinez-Valpuesta} et~al.}{2006}]{Martinez-Valpuesta.etal.2006}
{Martinez-Valpuesta} I.,  {Shlosman} I.,   {Heller} C.,  2006, \mn@doi [\apj]
  {10.1086/498338}, \href
  {https://ui.adsabs.harvard.edu/abs/2006ApJ...637..214M} {637, 214}

\bibitem[\protect\citeauthoryear{{Masters} et~al.,}{{Masters}
  et~al.}{2011}]{Masters.etal.2011}
{Masters} K.~L.,  et~al., 2011, \mn@doi [\mnras]
  {10.1111/j.1365-2966.2010.17834.x}, \href
  {https://ui.adsabs.harvard.edu/abs/2011MNRAS.411.2026M} {411, 2026}

\bibitem[\protect\citeauthoryear{{Navarro}, {Frenk}  \& {White}}{{Navarro}
  et~al.}{1996}]{NFW1996}
{Navarro} J.~F.,  {Frenk} C.~S.,   {White} S. D.~M.,  1996, \mn@doi [\apj]
  {10.1086/177173}, \href
  {https://ui.adsabs.harvard.edu/abs/1996ApJ...462..563N} {462, 563}

\bibitem[\protect\citeauthoryear{{O'Brien}, {Freeman}  \& {van der
  Kruit}}{{O'Brien} et~al.}{2010}]{O'Brien.etal.2010}
{O'Brien} J.~C.,  {Freeman} K.~C.,   {van der Kruit} P.~C.,  2010, \mn@doi
  [\aap] {10.1051/0004-6361/200912568}, \href
  {https://ui.adsabs.harvard.edu/abs/2010A&A...515A..63O} {515, A63}

\bibitem[\protect\citeauthoryear{{Quillen}, {Minchev}, {Sharma}, {Qin}  \& {Di
  Matteo}}{{Quillen} et~al.}{2014}]{Quillen.etal.2014}
{Quillen} A.~C.,  {Minchev} I.,  {Sharma} S.,  {Qin} Y.-J.,   {Di Matteo} P.,
  2014, \mn@doi [\mnras] {10.1093/mnras/stt1972}, \href
  {https://ui.adsabs.harvard.edu/abs/2014MNRAS.437.1284Q} {437, 1284}

\bibitem[\protect\citeauthoryear{{Raha}, {Sellwood}, {James}  \& {Kahn}}{{Raha}
  et~al.}{1991}]{Raha.etal.1991}
{Raha} N.,  {Sellwood} J.~A.,  {James} R.~A.,   {Kahn} F.~D.,  1991, \mn@doi
  [\nat] {10.1038/352411a0}, \href
  {https://ui.adsabs.harvard.edu/abs/1991Natur.352..411R} {352, 411}

\bibitem[\protect\citeauthoryear{{Saha} \& {Naab}}{{Saha} \&
  {Naab}}{2013}]{Saha.Naab.2013}
{Saha} K.,  {Naab} T.,  2013, \mn@doi [\mnras] {10.1093/mnras/stt1088}, \href
  {https://ui.adsabs.harvard.edu/abs/2013MNRAS.434.1287S} {434, 1287}

\bibitem[\protect\citeauthoryear{{Saha}, {Martinez-Valpuesta}  \&
  {Gerhard}}{{Saha} et~al.}{2012}]{Saha.etal.2012}
{Saha} K.,  {Martinez-Valpuesta} I.,   {Gerhard} O.,  2012, \mn@doi [\mnras]
  {10.1111/j.1365-2966.2011.20307.x}, \href
  {https://ui.adsabs.harvard.edu/abs/2012MNRAS.421..333S} {421, 333}

\bibitem[\protect\citeauthoryear{{Saha}, {Gerhard}  \&
  {Martinez-Valpuesta}}{{Saha} et~al.}{2016}]{saha.etal.2016}
{Saha} K.,  {Gerhard} O.,   {Martinez-Valpuesta} I.,  2016, \mn@doi [\aap]
  {10.1051/0004-6361/201527566}, \href
  {https://ui.adsabs.harvard.edu/abs/2016A&A...588A..42S} {588, A42}

\bibitem[\protect\citeauthoryear{{Savitzky} \& {Golay}}{{Savitzky} \&
  {Golay}}{1964}]{Savgol.1964}
{Savitzky} A.,  {Golay} M.~J.~E.,  1964, Analytical Chemistry, \href
  {https://ui.adsabs.harvard.edu/abs/1964AnaCh..36.1627S} {36, 1627}

\bibitem[\protect\citeauthoryear{{Sellwood}}{{Sellwood}}{1980}]{Sellwood.1980}
{Sellwood} J.~A.,  1980, \aap, \href
  {https://ui.adsabs.harvard.edu/abs/1980A&A....89..296S} {89, 296}

\bibitem[\protect\citeauthoryear{{Sellwood} \& {Gerhard}}{{Sellwood} \&
  {Gerhard}}{2020}]{Sellwood.Gerhard.2020}
{Sellwood} J.~A.,  {Gerhard} O.,  2020, \mn@doi [\mnras]
  {10.1093/mnras/staa1336}, \href
  {https://ui.adsabs.harvard.edu/abs/2020MNRAS.495.3175S} {495, 3175}

\bibitem[\protect\citeauthoryear{{Sellwood} \& {Merritt}}{{Sellwood} \&
  {Merritt}}{1994}]{Sellwood.Merritt.1994}
{Sellwood} J.~A.,  {Merritt} D.,  1994, \mn@doi [\apj] {10.1086/174004}, \href
  {https://ui.adsabs.harvard.edu/abs/1994ApJ...425..530S} {425, 530}

\bibitem[\protect\citeauthoryear{{Smirnov} \& {Sotnikova}}{{Smirnov} \&
  {Sotnikova}}{2019}]{Smirnov.2019}
{Smirnov} A.~A.,  {Sotnikova} N.~Y.,  2019, \mn@doi [\mnras]
  {10.1093/mnras/stz546}, \href
  {https://ui.adsabs.harvard.edu/abs/2019MNRAS.485.1900S} {485, 1900}

\bibitem[\protect\citeauthoryear{{Springel}}{{Springel}}{2005a}]{Springal2005man}
{Springel} V.,  2005a, {User guide for GADGET-2}.
\url {https://wwwmpa.mpa-garching.mpg.de/gadget/users-guide.pdf}

\bibitem[\protect\citeauthoryear{{Springel}}{{Springel}}{2005b}]{Springel2005}
{Springel} V.,  2005b, \mn@doi [\mnras] {10.1111/j.1365-2966.2005.09655.x},
  \href {https://ui.adsabs.harvard.edu/abs/2005MNRAS.364.1105S} {364, 1105}

\bibitem[\protect\citeauthoryear{{Springel}, {Yoshida}  \& {White}}{{Springel}
  et~al.}{2001}]{Springel2001}
{Springel} V.,  {Yoshida} N.,   {White} S. D.~M.,  2001, \mn@doi [\na]
  {10.1016/S1384-1076(01)00042-2}, \href
  {https://ui.adsabs.harvard.edu/abs/2001NewA....6...79S} {6, 79}

\bibitem[\protect\citeauthoryear{{Weinberg} \& {Katz}}{{Weinberg} \&
  {Katz}}{2007}]{Weinberg.etal.2007}
{Weinberg} M.~D.,  {Katz} N.,  2007, \mn@doi [\mnras]
  {10.1111/j.1365-2966.2006.11307.x}, \href
  {https://ui.adsabs.harvard.edu/abs/2007MNRAS.375..460W} {375, 460}

\bibitem[\protect\citeauthoryear{{Xiang} et~al.,}{{Xiang}
  et~al.}{2021}]{Xiang.etal.2021}
{Xiang} K.~M.,  et~al., 2021, \mn@doi [\apj] {10.3847/1538-4357/abdab5}, \href
  {https://ui.adsabs.harvard.edu/abs/2021ApJ...909..125X} {909, 125}

\bibitem[\protect\citeauthoryear{{Yurin} \& {Springel}}{{Yurin} \&
  {Springel}}{2014}]{Yurin2014}
{Yurin} D.,  {Springel} V.,  2014, \mn@doi [\mnras] {10.1093/mnras/stu1421},
  \href {https://ui.adsabs.harvard.edu/abs/2014MNRAS.444...62Y} {444, 62}

\makeatother
\end{thebibliography}






\bsp	
\label{lastpage}
\end{document}